\documentclass[aps,pra,amsmath,amssymb,reprint]{revtex4-2}

\usepackage[utf8]{inputenc}
\usepackage{amsmath}
\usepackage{amssymb}
\usepackage{xspace}
\usepackage{graphicx}
\usepackage{dcolumn}
\usepackage{braket}
\usepackage{float}
\usepackage{color}
\usepackage{xcolor}

\usepackage{siunitx}

\newcommand{\nop}[1]{n_{#1}}
\newcommand{\cop}[1]{a^{#1}} 
\newcommand{\aop}[1]{a_{#1}} 
\newcommand{\sqop}[2]{a_{#2}^{#1}}
\newcommand{\comm}[2]{[{#1},{#2}]}
\newcommand{\acomm}[2]{\{{#1},{#2}\}}

\usepackage[capitalize]{cleveref}

\DeclareUnicodeCharacter{2212}{-}
\DeclareUnicodeCharacter{0301}{\'{e}}
\DeclareUnicodeCharacter{0308}{\"}
\DeclareUnicodeCharacter{03C3}{$\sigma$}
\DeclareUnicodeCharacter{010C}{\v{C}}
\DeclareUnicodeCharacter{00ED}{\'{i}}
\DeclareUnicodeCharacter{017E}{\v{z}}

\begin{document}

\title{Exact closed-form unitary transformations of fermionic operators}

\author{Francesco A. Evangelista}
\email{francesco.evangelista@emory.edu}
\author{Ilias Magoulas}
\affiliation{Department of Chemistry and Cherry Emerson Center for Scientific Computation, Emory University, Atlanta, Georgia 30322, USA}

\date{\today}

\begin{abstract}
Unitary transformations play a fundamental role in many-body physics, and except for special cases, they are not expressible in closed form.
We present closed-form expressions for unitary transformations generated by a single fermionic operator for Hermitian and anti-Hermitian generators.
We demonstrate the usefulness of these expressions in formal analyses of unitary transformations and numerical applications to Hamiltonian downfolding in quantum computing and Heisenberg dynamics.
This work paves the way for new analytical treatments of unitary transformations and numerical many-body methods for fermions.
\end{abstract}

\maketitle

\section{Introduction}
Hamiltonian transformations are a powerful tool for expressing quantum many-body problems in an equivalent form that is analytically solvable or easier to solve numerically due to a reduction of the relevant degrees of freedom.
In particular, similarity transformations play a foundational role in analytic treatments of many-body interactions \cite{Vleck.1929.10.1103/physrev.33.467,Bogoljubov.1958.10.1007/BF02745585,Valatin.1958.10.1007/BF02745589,Schrieffer.1966.10.1103/physrev.149.491,Bravyi.2011.10.1016/j.aop.2011.06.004,wagner1986unitary,Ashida.2018.10.1103/physrevlett.121.026805} and various quantum many-body methods, including coupled-cluster \cite{Coester.1958.10.1016/0029-5582(58)90280-3,Coester.1960.10.1016/0029-5582(60)90140-1,Cizek.1966.10.1063/1.1727484,Cizek.1969.10.1002/9780470143599.ch2,Bartlett.1989.10.1016/s0009-2614(89)87372-5,doi:https://doi.org/10.1002/9780470141694.ch1,Bartlett.2007.10.1103/revmodphys.79.291} and canonical transformation theories \cite{White.2002.10.1063/1.1508370,Yanai.2006.10.1063/1.2196410}, renormalization group methods \cite{Glazek.1993.10.1103/physrevd.48.5863,Wegner.1994.10.1002/andp.19945060203,Tsukiyama.2011.10.1103/physrevlett.106.222502,Evangelista.2014.10.1063/1.4890660}, and reduced density matrix approaches \cite{Mazziotti.2012.10.1021/cr2000493}.
A common goal of these approaches is to bring a general many-body Hamiltonian ($\mathcal{H}$) to a full or partially diagonal form ($\bar{\mathcal{H}}$) via a similarity transformation expressed in terms of a generator $S$:
\begin{equation}
\label{eq:simtrans}
	\mathcal{H} \mapsto \bar{\mathcal{H}} = e^{-S} \mathcal{H} e^{S}.
\end{equation}
In a few cases, this similarity transformation can be evaluated analytically.
Examples include the Bogoliubov transformation used to diagonalize model Hamiltonians that phenomenologically describe superfluidity (bosons) and superconductivity (fermions) \cite{Valatin.1958.10.1007/BF02745589,Bogoljubov.1958.10.1007/BF02745585}, and  Pauli rotations of Pauli operators in quantum computing \cite{gottesman1998heisenbergrepresentationquantumcomputers,Begusic.2024.10.1126/sciadv.adk4321}.

Focusing on the case of fermionic operators, on the numerical side, the similarity-transformed Hamiltonian of conventional coupled-cluster theory can be evaluated exactly, albeit with some effort. In this case, $S$ is a particle--hole substitution operator, promoting electrons from occupied orbitals (hole states) to the unoccupied ones (particle states) in a reference Slater determinant that serves as the Fermi vacuum.
For this choice of $S$, closed-form expressions for $\bar{\mathcal{H}}$ can be evaluated using a lemma of the Baker--Campbell--Hausdorff (BCH) formula, which expresses the similarity transformation of an operator $O$ as an infinite commutator series:
\begin{equation}
\label{eq:bch}
	e^{-S} O e^{S} = O + [O,S] + \frac{1}{2!}[[O,S],S] + \ldots.
\end{equation}
With the choice of operators made in traditional coupled-cluster theory, this series truncates after the $2k$-fold commutator term for a Hamiltonian with up to $k$-body interactions.
From a diagrammatic perspective, and considering that commutators represent connected quantities, $\bar{\mathcal{H}}$ has a finite expansion because the Hamiltonian vertex must be connected to all $S$ vertices, and the $S$ vertices cannot connect with each other. 

However, in general, similarity transformations cannot be written in closed form.
In numerical applications, this means that $\bar{\mathcal{H}}$ must be approximated.
For example, in unitary coupled cluster and related formalisms \cite{Kutzelnigg.1982.10.1063/1.444231,Bartlett.1989.10.1016/s0009-2614(89)87372-5,White.2002.10.1063/1.1508370,Yanai.2006.10.1063/1.2196410,Sokolov.2014.10.1063/1.4892946,Filip.2020.10.1063/5.0026141,Misiewicz.2021.10.1063/5.0076888,Liu.2022.10.1021/acs.jctc.1c01210,Grimsley.2022.10.1021/acs.jctc.2c00751}, the BCH expansion of $\bar{\mathcal{H}}$ does not terminate, as one can connect $S$ vertices to both $S$ and $\mathcal{H}$.
Consequently, $\bar{\mathcal{H}}$ is typically approximated by truncating the commutator expansion using perturbative arguments  \cite{Watts.1989.10.1016/0009-2614(89)87262-8}, by neglecting higher-body terms in the commutator expansion \cite{Yanai.2006.10.1063/1.2196410,Evangelista.2012.10.1016/j.chemphys.2011.08.006}, or using infinite-order approximations to the exponential transformation \cite{Kutzelnigg.1991.10.1007/bf01117418,Taube.2006.10.1002/qua.21198,Grimsley.2022.10.1021/acs.jctc.2c00751}.
The past decade has seen a resurgence of interest in unitary formalism both from the perspective of classical and quantum computing.
Similarity renormalization group methods that perform unitary transformations of the Hamiltonian have been developed in nuclear \cite{Tsukiyama.2011.10.1103/physrevlett.106.222502,Hergert.2016.10.1016/j.physrep.2015.12.007}
 and electronic structure theory \cite{White.2002.10.1063/1.1508370,Evangelista.2014.10.1063/1.4890660,Li.2016.10.1063/1.4947218,Li.2019.10.1146/annurev-physchem-042018-052416}.
Unitary downfolding/dimensionality reduction methods have been proposed to reduce the cost of classical and quantum computations \cite{Bauman.2019.10.1063/1.5094643,Takeshita.2020.10.1103/PhysRevX.10.011004,Kowalski.2024.10.1063/5.0207534,Sahinoglu.2021.10.1038/s41534-021-00451-w,Huang.2023.10.1103/prxquantum.4.020313,Kowalski.2023.10.1103/physrevlett.131.200601}.
Quantum computers open new avenues for the implementation of factorized unitary coupled-cluster and related many-body trial states based on products of fermionic rotations \cite{Peruzzo.2014.10.1038/ncomms5213,Wecker.2015.10.1103/physreva.92.042303,Lee.2019.10.1021/acs.jctc.8b01004,Evangelista.2019.10.1063/1.5133059,Lang.2021.10.1021/acs.jctc.0c00170,Anselmetti.2021.10.1088/1367-2630/ac2cb3,Anand.2022.10.1039/d1cs00932j,Henderson.2023.10.1080/00268976.2023.2254857,Burton.2023.10.1038/s41534-023-00744-2,Burton.2024.10.1103/PhysRevResearch.6.023300}.
Quantum-inspired classical algorithms based on unitary transformations of Pauli operators have also been proposed to block-diagonalize Hamiltonians \cite{Ryabinkin.2020.10.1021/acs.jctc.9b01084}, to perform Heisenberg dynamics \cite{Begusic.2024.10.1126/sciadv.adk4321}, and evaluate expectation values of operators \cite{begusic2023simulatingquantumcircuitexpectation}.

In this work, we examine the special case of unitary transformations generated by a \textit{single} product of fermionic creation and annihilation operators $T$, considering both the case of anti-Hermitian $S = \theta A = \theta (T - T^\dagger)$ and Hermitian $S = -i\theta H = -i \theta (T + T^\dagger)$ generators, where  $\theta$ is a real parameter.
For convenience, we refer to these transformations as fermionic rotations.
We show that for this class of transformations, it is possible to derive closed-form expressions for the transformed operator and the corresponding operator flow.
These equations are expressed in terms of up to doubly-nested commutators, independently from the rank of the operators $O$ and $T$ (where we define the rank as half the number of creation plus the annihilation operators).
Once combined with Trotterization, our result enables the realization of general unitary transformations with numerical control.
We demonstrate this point via a numerical implementation of fermionic rotations, which we use to realize an adaptive block-diagonalization of molecular Hamiltonians and to implement Heisenberg dynamics in the Hubbard model.

\begin{figure*}[ht!]
 \includegraphics[width=7in]{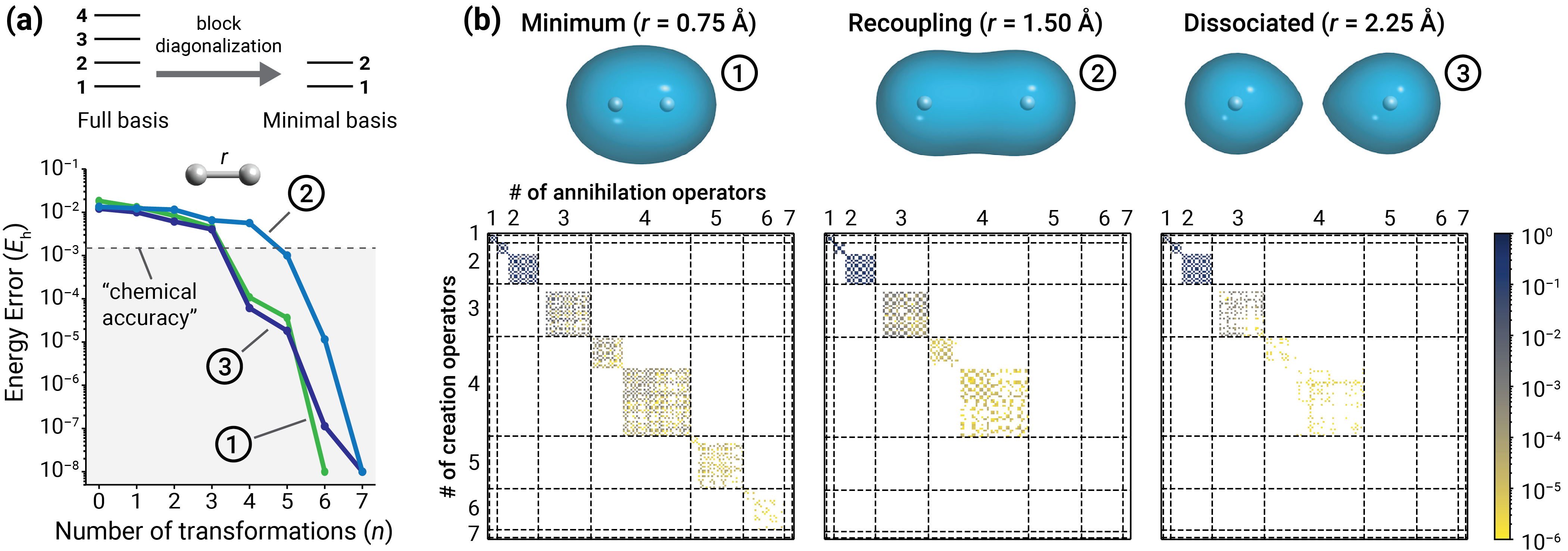}
  \caption{Application of unitary downfolding to the H\textsubscript{2} molecule at three representative geometries.
  (a) Convergence of the energy error (in hartree, $E_\mathrm{h}$) with respect to exact diagonalization vs.\ number of operators in the sequence of transformations [see \cref{eq:simtrantrott}].
  (b) The magnitude of the components of the similarity-transformed Hamiltonian ($\bar{\mathcal{H}}$) at the converged transformation for the three geometries.
The tensor elements of $\bar{\mathcal{H}}$ are represented as a matrix, with rows (columns) corresponding to products of creation (annihilation) operators, both grouped according to the product length (ranging from 0 to 8). In this example, a full optimization of the parameters was performed after selecting a new operator.}
  \label{fig:downfolding}
\end{figure*}

\section{Theory}

We begin by introducing the notation used in this work.
We consider products of fermionic operators with creation operators ($\cop{p}$) arranged to the left of annihilation operators ($\aop{p}$).
Such products are represented with the following compact notation
\begin{equation}
\cop{p_1} \cop{p_2} \cdots \cop{p_n} \aop{q_m} \cdots \aop{q_2} \aop{q_1} = \sqop{p_1 p_2 \cdots p_n}{q_1 q_2 \cdots q_m},
\end{equation}
where we assume that all operator indices are sorted in increasing order (e.g., $p_1 < p_2 < \ldots < p_n$, $q_1 < q_2 < \ldots < q_m$).
Number operators are indicated with the notation $\nop{p} = \cop{p} \aop{p} = \sqop{p}{p}$.

\subsection{Closed-form transformations: Anti-Hermitian case}
Our main result concerns how a fermionic operator product
\begin{equation}
\label{eq:o_def}
O = a_{q_1 q_2 \cdots}^{p_1 p_2 \cdots},
\end{equation}
transforms under a unitary transformation generated by the anti-Hermitian operator $A$:
\begin{equation}
\label{eq:similarity_transformation}
O \mapsto \bar{O} = e^{-\theta A} O e^{\theta A},
\end{equation}
where $\theta$ is a real parameter and $A = T - T^\dagger$ is a many-body generator defined by the fermionic operator product $T$.
The case of a Hermitian operator times the imaginary unit will be discussed below.
We consider the general case of $T$, which consists of second quantized operators with distinct indices ($\sqop{s_1 s_2 \cdots}{t_1 t_2 \cdots}$) and optionally a number operator component ($\sqop{r_1 r_2 \cdots}{r_1 r_2 \cdots} = \nop{r_1} \nop{r_2} \cdots$):
\begin{equation}
\label{eq:t_def}
T = \sqop{s_1 s_2 \cdots r_1 r_2 \cdots}{t_1 t_2 \cdots r_1 r_2 \cdots}
= \nop{r_1} \nop{r_2} \cdots \sqop{s_1 s_2 \cdots}{t_1 t_2 \cdots}.
\end{equation}
All indices that define $T$ refer to generic spinorbitals, not restricted to the particle--hole picture.
Note that if $T$ is simply a product of number operators, then $T^\dagger = T$ and $A = 0$.
Furthermore, due to the properties of the fermionic algebra, if any two upper or lower indices are repeated, then $A = 0$ as well. In these cases, the transformation is trivial.

Next, we examine the property of commutators to expose a recurrence relationship that results from our choice for $A$.
As shown in \cref{app:recursion}, $A$ satisfies $A^3 = -A$ and the following closure relationship:
\begin{equation}
\comm{\comm{\comm{O}{A}}{A}}{A} = - \comm{O}{A}- 3 A\comm{O}{A}A.
\end{equation}
As further proven in the \cref{app:commutator}, based on the structures of $T$ and $O$, the quantity $A\comm{O}{A}A$ can take two values, either $\comm{O}{A}$ or 0.
Using this result, we may express the triply nested commutator as:
\begin{equation}
\label{eq:c3OA}
\comm{\comm{\comm{O}{A}}{A}}{A} = -\alpha \comm{O}{A},
\end{equation}
where $\alpha = 1$ (when $A\comm{O}{A}A = 0$) or $\alpha = 4$ (when $A\comm{O}{A}A = \comm{O}{A}$).
Taking repeated commutators of \cref{eq:c3OA} with $A$, we arrive at the following recursive conditions for the odd and even commutators of $O$ and $A$:
\begin{align}
\label{eq:recursive_condition1}
[[\ldots[O,\underbrace{A],A],\ldots],A]}_{2k+1} & = (-\alpha)^{k} \comm{O}{A} & (k \geq 0),\\
\label{eq:recursive_condition2}
[[\ldots[O,\underbrace{A],A],\ldots],A]}_{2k} & = (-\alpha)^{k-1} \comm{\comm{O}{A}}{A} & (k \geq 1).
\end{align}
\Cref{eq:recursive_condition1,eq:recursive_condition2} may be used to re-sum the BCH series for the transformed operator and obtain the following general expression:
\begin{equation}
\label{eq:expmAOexpA}
\bar{O} = O + \frac{\sin(\sqrt{\alpha} \theta)}{\sqrt{\alpha}}\comm{O}{A} +\frac{1-\cos(\sqrt{\alpha} \theta)}{\alpha}\comm{\comm{O}{A}}{A}.
\end{equation}
\Cref{eq:expmAOexpA} reduces to the following when $\alpha = 1$:
\begin{equation}
\label{eq:sto_case1}
\bar{O} = O + \sin\theta \comm{O}{A} 
+(1 - \cos\theta) \comm{\comm{O}{A}}{A},
\end{equation}
while when $\alpha = 4$ it leads to:
\begin{equation}
\label{eq:sto_case2}
\bar{O} = O + \frac{1}{2} \sin (2\theta) \, \comm{O}{A} 
+\frac{1}{2}\sin^2\theta \,\comm{\comm{O}{A}}{A}.
\end{equation}
Using the closed-form expression for the unitary transformation, we may also evaluate the operator flow as a function of $\theta$:
\begin{equation}
\label{eq:sto_flow}
\frac{d}{d\theta}\bar{O} = \cos(\sqrt{\alpha} \theta)\comm{O}{A} +\frac{1}{\sqrt{\alpha}} \sin(\sqrt{\alpha} \theta)\comm{\comm{O}{A}}{A}.
\end{equation}
To the best of our knowledge, \cref{eq:sto_case1,eq:sto_case2} are the first example of equations expressing a general class of fermionic unitary transformations in closed form.
Previous work by Wagner \cite{wagner1986unitary}, reported closed-form expressions for multiparticle unitary transformations for a few special cases, with the most complex one corresponding to the operators $O = \sqop{p_1 p_2}{q_1 p_2}$ and $T = \sqop{p_1}{q_1}$.

An alternative strategy for deriving \cref{eq:sto_case1,eq:sto_case2} is starting from the closed-form expression for $\exp(\theta A)$:
\begin{equation}
\label{eq:expA}
\exp(\theta A) = 1 + \sin \theta A + (1 - \cos\theta) A^2.
\end{equation}
This expression enables the efficient evaluation of states built from a product of fermionic unitary operators applied to an uncorrelated state, and it finds applications in classical algorithms \cite{Filip.2020.10.1063/5.0026141} and in the efficient emulation of quantum  algorithms \cite{Evangelista.2019.10.1063/1.5133059,Chen.2021.10.1021/acs.jctc.0c01052,Rubin.2021.10.22331/q-2021-10-27-568,Xu.2023.10.3390/sym15071429}.
A derivation of the operator transformation $\exp(-\theta A) O \exp(\theta A)$ starting from \cref{eq:expA} does not directly yield manifestly connected expressions such as \cref{eq:sto_case1,eq:sto_case2} (though they may be shown to be equivalent to those derived here).
Crucially, even in this alternative derivation, proving that \cref{eq:c3OA} holds for fermionic rotations is essential to express the final equations in terms of only single and double commutators of $O$ and $A$.
Note that our derivation does not assume a particle--hole restriction and allows for different numbers of creation and annihilation operators as well as repeated indices in $O$ and $T$.

To illustrate the usefulness of \cref{eq:sto_case1,eq:sto_case2}, we reproduce a textbook case \cite{wagner1986unitary} of unitary transformations of one-body fermionic operators.
We begin by considering the following Hamiltonian
$\mathcal{H} = \alpha \nop{p} + \beta \nop{q} + \gamma (\sqop{p}{q}  + \sqop{q}{p})$
involving distinct spinorbitals $\psi_p$ and $\psi_q$ and consider rotations generated by $T=\sqop{p}{q}$ with the purpose of eliminating the off-diagonal term $\sqop{p}{q} + \sqop{q}{p}$.
For each term $O$ in the Hamiltonian, we begin by evaluating $[O,A]$ and $A[O,A]A$ to determine the transformation case. One can show that for all terms $A[O,A]A = [O,A]$, so we use \cref{eq:sto_case1}. A straightforward application of this equation yields:

\begin{equation}
	\begin{split}
e^{-\theta (\sqop{p}{q}-\sqop{q}{p})}  \nop{p} e^{\theta (\sqop{p}{q}-\sqop{q}{p})} 
 = \; &
\cos^2 \theta \; \nop{p} +\sin^2 \theta \nop{q} \\& + \frac{1}{2}\sin (2\theta)  (\sqop{p}{q} + \sqop{q}{p}),
	\end{split}
\end{equation}
\begin{equation}
	\begin{split}
e^{-\theta (\sqop{p}{q}-\sqop{q}{p})} \nop{q} e^{\theta (\sqop{p}{q}-\sqop{q}{p})} 
 = \; &
\cos^2 \theta \; \nop{q} +\sin^2 \theta \nop{p} \\& - \frac{1}{2}\sin (2\theta)  (\sqop{p}{q} + \sqop{q}{p}),
	\end{split}
\end{equation}
while the off-diagonal term transforms as
\begin{equation}
	\begin{split}
e^{-\theta (\sqop{p}{q}-\sqop{q}{p})} (\sqop{p}{q}  + \sqop{q}{p})e^{\theta (\sqop{p}{q}-\sqop{q}{p})} 
 = \; & \cos (2\theta)  (\sqop{p}{q} +\sqop{q}{p}) \\
 &+ \sin (2\theta)(\nop{q} - \nop{p}).
	\end{split}
\end{equation}
Using these expressions to transform the Hamiltonian and isolating the term proportional to $\sqop{p}{q} +\sqop{q}{p}$, we find that the coupling can be eliminated when $\theta$ is a solution of the following equation:
\begin{equation}	
\frac{1}{2} \sin (2\theta) (\alpha - \beta) + \gamma \cos (2\theta) = 0,
\end{equation}
or equivalently $\tan (2\theta) = 2 \gamma / (\beta - \alpha)$.

As a more complex example, we consider the anti-Hermitian transformation of a two-body operator $O = \sqop{p_1 p_2}{q_1 q_2}$ by the two-body generator $T = \sqop{p_1 s_1}{q_1 t_1}$.
In this case, we have $A[O,A]A = 0$ [$\alpha = 1$ case, \cref{eq:sto_case1}], and the transformed operator is equal to
\begin{equation}
\begin{split}
\bar{O} = \; & \sqop{p_1 p_2}{q_1 q_2} + \sin \theta \sqop{p_2 t_1}{q_2 s_1} (\nop{q_1} - \nop{p_1}) \\
&+ (1-\cos \theta) \sqop{p_1 p_2}{q_1 q_2} (2 \nop{s_1}\nop{t_1} - \nop{s_1} - \nop{t_1}).	
\end{split}
\end{equation}
In contrast to the previous example, this transformation increases the operator rank, introducing three- and four-body induced interactions in the form of products of the original operator times number operators.
These closed-form equations can similarly be derived for any combination of operators $O$ and $T$, providing a systematic approach to examining the effects of fermionic rotations.

\subsection{Closed-form transformations: Hermitian case}
Closed-form expressions may also be derived for transformations generated by the Hermitian operator ($H =  T + T^\dagger$) times the imaginary unit, using the recursion relationship:
\begin{equation}
\label{eq:c3OS}
\comm{\comm{\comm{O}{H}}{H}}{H} = \beta \comm{O}{H},
\end{equation}
where $\beta$ can be shown to be equal to 1 or 4 depending on the structure of $T$ and $O$.
Carrying out a derivation analogous to the anti-Hermitian case, the following formulas for a transformed operator may be derived for the case $\beta = 1$:
\begin{equation}
\label{eq:hermitian_1}
	e^{i \theta H}  O e^{-i \theta H} 
=  O -i\sin\theta [O,H] + (\cos\theta - 1) [[O,H],H],
\end{equation}
and for the case $\beta = 4$:
\begin{equation}
\label{eq:hermitian_2}
	e^{i \theta H}  O e^{-i \theta H} 
=  O -\frac{i}{2}\sin(2\theta) [O,H] - \frac{1}{2}\sin^2(\theta) [[O,H],H].
\end{equation}
In contrast to the anti-Hermitian case where a number operator ($T = \nop{r_1} \nop{r_2} \cdots$) results in a trivial transformation, here it falls under the case of \cref{eq:hermitian_2} (since for this case, $\beta = 4$). 

To verify the correctness of our closed-form expressions and generate numerical results, we have implemented them both using a symbolic manipulator \cite{Evangelista.2022.10.1063/5.0097858} and codes that numerically implement the algebra of fermionic operators \cite{Evangelista.2024.10.1063/5.0216512}.

\subsection{Comparison with Pauli rotations}
It is interesting to compare the fermionic transformations derived here with Pauli rotations of a  Pauli operator $O$ (tensor product of Pauli matrices) \cite{begusic2023simulatingquantumcircuitexpectation}:
\begin{equation}
\label{eq:Pauli}
e^{i \theta P / 2} O e^{-i \theta P / 2} =
\begin{cases}
O, & \comm{P}{O} = 0,\\
O \cos \theta + i \sin \theta PO, & \acomm{P}{O}= 0,
\end{cases}
\end{equation}
where $P$ is a Pauli operator and $\acomm{P}{O} = PO + OP$.
The simpler structure of this transformation stems from the Pauli operator algebra, which implies $P^2=1$ and that two Pauli operators either commute or anticommute.
Therefore, in the case of Pauli operators, only knowledge of the product $PO$ is required, instead of the single and double commutators as in the fermionic case.
Although one may, in principle, map fermionic operators to Pauli operators and perform transformations via \cref{eq:Pauli} \cite{Ryabinkin.2020.10.1021/acs.jctc.9b01084}, working in the fermionic representation has two advantages: 1) it is more compact as the mapping generally leads to a linear combination of commuting Pauli operators with the number of terms scaling exponentially with respect to operator rank, and 2) it is straightforward to preserve symmetries like particle number, point group, and spin projection.

\section{Numerical Results}
\subsection{Hamiltonian downfolding}

\begin{figure}[ht!]
 \includegraphics[width=3.375in]{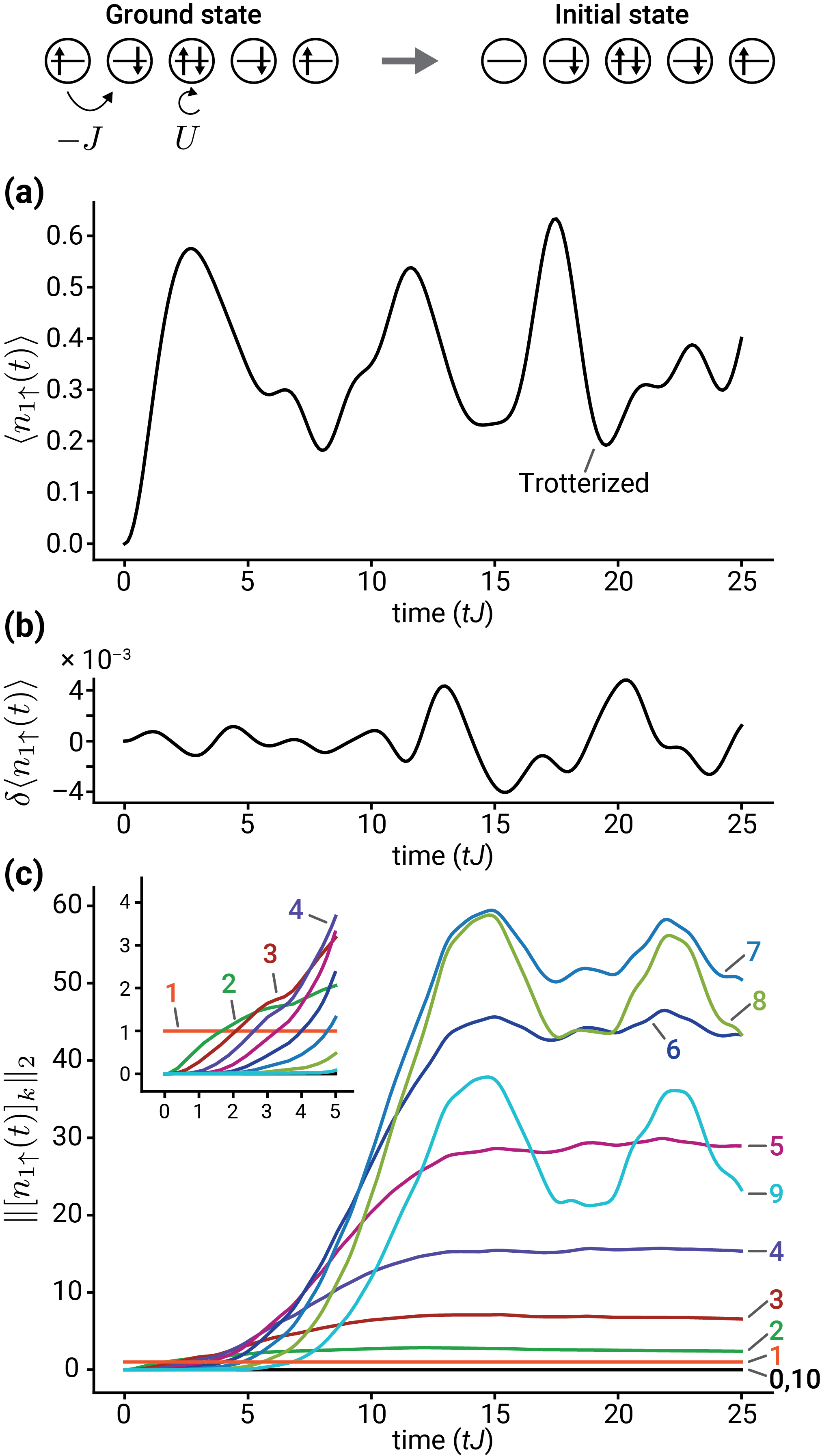}
  \caption{Dynamics following sudden ionization of a 5-site Hubbard model with open boundaries with hopping parameter $J = 1$ and on-site Coulomb integral $U = 1$.
  To simulate the sudden ionization of a spin-up electron on site 1, the initial state ($t = 0$) is obtained from the ground state with 3 spin-up and 3 spin-down electrons by application of the annihilation operator $a_{1\uparrow}$ followed by normalization.
  (a) Expectation value of the number of spin-up electrons on site 1, $\langle n_{1\uparrow}(t) \rangle$, as a function of time (in units of $tJ$) for the Trotterized dynamics evaluated via \cref{eq:hermitian_1,eq:hermitian_2} (second-order Trotter approximation with 200 steps).
  (b) Deviation of $\langle n_{1\uparrow}(t) \rangle$  computed with the Trotterized dynamics from the results of exact dynamics, $\delta \langle n_{1\uparrow}(t) \rangle$.
  (c) Euclidean norm of the $k$-body component of the number operator $n_{1\uparrow}(t) = a^\dagger_{1\uparrow}(t) a_{1\uparrow}(t)$, $\| [n_{1\uparrow}(t) ]_k\|_2$, in the Heisenberg representation (ranging from 0 to 10).
  The inset shows the Euclidean norm for $tJ$ up to 5.}
  \label{fig:dynamics}
\end{figure}

\begin{figure}[ht!]
 \includegraphics[width=3.375in]{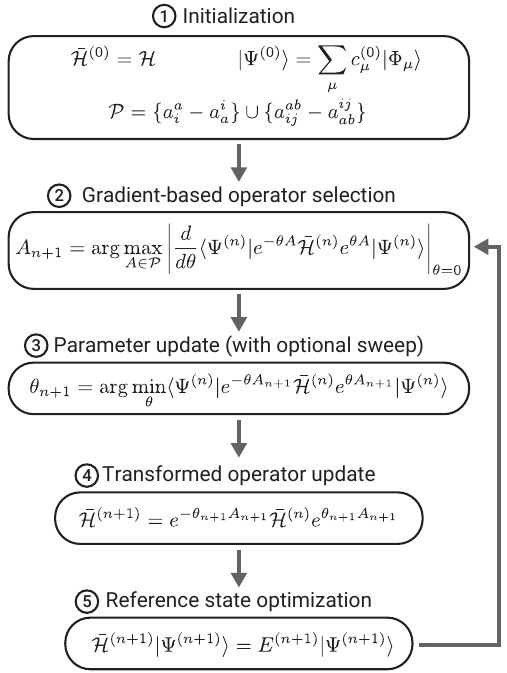}
  \caption{Diagram illustrating the Hamiltonian downfolding process.
  The operator pool ($\mathcal{P}$) contains one- and two-body operators that promote electrons from the occupied ($i,j$) to the unoccupied ($a,b$) orbitals of $\Psi$.
  The next operator to be added to the transformation ($A_{n+1}$) is selected in Step 2 based on the absolute value of the energy gradient. In Step 3, the parameter corresponding to $A_{n+1}$ is variationally optimized, and optionally, a single optimization sweep is performed in which all parameters in the transformation are individually optimized.}
  \label{fig:downfolding_algo}
\end{figure}

Next, we illustrate, using a pilot implementation, how the closed-form fermionic transformations presented in this work may be used in numerical applications.
In the first example, we demonstrate how our closed-form expression could be used to perform an exact canonical transformation of Hamiltonians, enabling a rigorous analysis of the importance of three- and higher-body induced interactions.
We consider the H\textsubscript{2} molecule with atomic distance $r$ in a Hartree--Fock basis built from a double-$\zeta$ basis consisting of atomic 1s and 2s orbitals, leading to a spinorbital basis composed of 8 functions $\{ \psi_{p\sigma}, p = 1,2,3,4, \sigma = \uparrow,\downarrow \}$ where $p$ = 1 and 2 are bonding/antibonding combinations of 1s-like orbitals, while $p$ = 3 and 4 are bonding/antibonding combinations of 2s-like orbitals.
We focus on three representative geometries along the dissociation curve, encompassing the weak correlations around the equilibrium geometry ($r = \SI{0.75}{\AA}$), the recoupling region ($r = \SI{1.25}{\AA}$), and the strong correlation regime near the dissociation limit ($r = \SI{2.25}{\AA}$).

We consider an adaptive downfolding procedure, which performs an iterative block-diagonalization of the Hamiltonian operator.
As illustrated in \cref{fig:downfolding_algo}, the second-quantized Hamiltonian is transformed to a block-diagonal form by eliminating the operator components that couple the spinorbitals $\psi_{1\sigma}$ and $\psi_{2\sigma}$ of the hydrogen molecule (dominated by bonding/antibonding combinations of 1s orbitals) from the remaining ones.
In this reduced space, due to symmetry, the singlet totally symmetric ground state is a combination of two determinants, $\ket{\Psi} = c_1 \ket{\psi_{1\uparrow} \psi_{1\downarrow}} + c_2 \ket{\psi_{2\uparrow} \psi_{2\downarrow}}$, where $c_1$ and $c_2$ are variational parameters and $\ket{\Psi}$ is normalized to one.
The Hamiltonian is transformed by a sequence of $n$ fermionic rotations $\{A_1, A_2, \ldots, A_n\}$: 
\begin{equation}
\label{eq:simtrantrott}
\bar{\mathcal{H}} = \prod_{j=n}^{1} e^{-\theta_j A_j} \mathcal{H} \prod_{j = 1}^{n} e^{\theta_j A_j},
\end{equation}
where the operators are selected from a pool $\mathcal{P}$ of single and double substitution operators that replace $\psi_{1\sigma}$ and $\psi_{2\sigma}$ with $\psi_{3\sigma}$ and $\psi_{4\sigma}$.
The sequence of operators is selected using a criterion based on the gradient of the expectation value of $\bar{\mathcal{H}}$ with respect to the ground state, $\braket{\Psi | \bar{\mathcal{H}} | \Psi}$, as done in the ADAPT-VQE algorithm \cite{Grimsley.2019.10.1038/s41467-019-10988-2}.
However, these operators are added in the reverse order used in ADAPT-VQE.
For each operator $A \in \mathcal{P}$, we evaluate its gradient as $\braket{\Psi | [\bar{\mathcal{H}},A] | \Psi}$ \cite{Grimsley.2019.10.1021/acs.jctc.9b01083}.
The operator with the largest absolute gradient is used to further transform the Hamiltonian.
After a new operator ($A_{n+1}$) is added to the transformation, the corresponding parameter ($\theta_{n+1}$) is optimized, and optionally, all other parameters are re-optimized. An improved $\ket{\Psi}$ is then recomputed by diagonalizing $\bar{\mathcal{H}}$ in the two-determinant subspace.
From a quantum chemistry perspective, this procedure is equivalent to performing a two-electron in two-orbital complete-active-space (CAS) computation that includes dynamical electron correlation corrections via downfolding
\cite{White.2002.10.1063/1.1508370,Yanai.2006.10.1063/1.2196410,Li.2016.10.1063/1.4947218,Hergert.2016.10.1016/j.physrep.2015.12.007,Bauman.2019.10.1063/1.5094643,Takeshita.2020.10.1103/PhysRevX.10.011004,Huang.2023.10.1103/prxquantum.4.020313,Kowalski.2023.10.1103/physrevlett.131.200601}.

As shown in panel (a) of \cref{fig:downfolding}, the energy error with respect to the exact ground-state energy converges to zero quickly with the number of operators added, independent of the strength of electron correlations.
At the same time, plots of the magnitude of the various operator components of $\bar{\mathcal{H}}$ show that the induced three- and higher-body interactions rapidly decay with increasing rank [panel (b) of \cref{fig:downfolding}].
The nature of the selected operator sequences and the magnitudes of the higher-body terms also show variations at different values of $r$.
For example, at $r = \SI{0.75}{\AA}$, the first two operators selected consist of pair substitutions ($a^{4 \uparrow 4 \downarrow}_{1 \uparrow 1 \downarrow}$, followed by $a^{3 \uparrow 3 \downarrow}_{1 \uparrow 1 \downarrow}$) which account for correlation effects outside the lowest two orbitals.
Instead, at $r = \SI{2.25}{\AA}$, the first two operators selected consist of single substitutions ($a^{3 \uparrow }_{1 \uparrow}$, followed by $a^{4 \uparrow}_{2 \uparrow}$), which account for orbital relaxation (hybridization) effects to correct deficiencies in the mean-field orbitals.
At the same time, as $r$ increases and the ground state becomes more entangled, the magnitude of the higher-body terms decreases.

\begin{table*}[ht!]
	\renewcommand{\arraystretch}{1.1}
	\caption{Comparison of unitary downfolding methods applied to the H$_2$ molecule for four representative bond distances. Energy errors (in m$E_\mathrm{h}$) are computed using Hartree--Fock orbitals expanded in a cc-pVTZ basis set (30 basis functions).
	Comparison of exact diagonalization in the full orbital space (FCI, absolute energies), exact diagonalization in the space of 1s, 2s orbitals (CASCI, not downfolded), double unitary coupled cluster (DUCC, from Ref.~\cite{Bauman.2019.10.1063/1.5094643}), quantum linearized driven similarity renormalization group with one- and two-body operators [2-QLDSRG(2), from Ref.~\cite{Huang.2023.10.1103/prxquantum.4.020313}], and adaptive downfolding based on exact unitary transformations. The number of parameters is the size of the diagonalization space plus the number of operators in the Hamiltonian transformation.}
\begin{ruledtabular}
\begin{tabular}{@{} lccrrrr @{}}			
Method & Size of downfolded basis & \# of parameters & 0.8 bohr & 1.4008 bohr & 4 bohr & 10 bohr \\
\hline
FCI (full space) & 30 & 166 & $-1.015729$ & $-1.172455$ & $-1.014872$ & $-0.999623$ \\
CASCI (active space) & 4 & 8 & 32.729 & 25.755 & 7.872 & 2.523 \\ 
DUCC     & 4 & 166 & 7.129 & 4.555 & $-0.328$ & $-1.977$ \\
2-QLDSRG(2) & 4 & 3096 & $-0.087$ & $-0.084$ & 3.794 & 0.126 \\
Adaptive downfolding (this work) & 4 & 12--138$^a$ & 0.909 & 0.353 & 0.730 & 0.184 \\
		\end{tabular}
	\end{ruledtabular}
	\label{tab:h2_benchmark}
\footnotesize{$^a$ In adaptive downfolding, the number of parameters varies with geometry: 138 at 0.8 bohr, 126 at 1.4008 bohr, 19 at 4 bohr, and 12 at 10 bohr.}	
\end{table*}

Next, we report a comparison with other many-body downfolding approaches that highlights the advantage of using exact unitary transformations.
In \cref{tab:h2_benchmark}, we consider an example from Ref.~\cite{Bauman.2019.10.1063/1.5094643}, where downfolding is applied to four representative geometries of the H$_2$ molecule using a medium-size basis consisting of 30 orbitals, reduced to 4 orbitals (combinations of 1s- and 2s-like orbitals) via downfolding.
The CASCI data in \cref{tab:h2_benchmark} show that diagonalization of the original Hamiltonian in the space of 4 orbitals results in large energy errors (2.5--32.7 m$E_\mathrm{h}$), significantly above the so-called \textit{chemical accuracy} target of ca. 1.6 m$E_\mathrm{h}$.
The double unitary coupled cluster (DUCC) \cite{Bauman.2019.10.1063/1.5094643} and the quantum driven similarity renormalization group [2-QLDSRG(2)] \cite{Huang.2023.10.1103/prxquantum.4.020313}---two classical unitary downfolding methods proposed for quantum computing applications---reduce the energy error at the cost of significant increases in the number of variational parameters.
In both cases, the number of parameters exceeds those required by exact diagonalization (166), suggesting that the accuracy of these methods is limited by the use of approximate transformations, resulting in loss of accuracy.
In contrast, adaptive downfolding based on exact unitary transformations achieves sub-m$E_\mathrm{h}$ errors across the four sampled geometries using a variable number of parameters.
At the compressed geometry ($r = 0.8$ bohr), this procedure produces the most complex transformation, requiring 138 parameters, whereas, in the dissociation limit ($r = 10$ bohr), it results in a much simpler transformation with only 12 parameters.

\subsection{Heisenberg dynamics}
Our next example focuses on simulating the Heisenberg dynamics via closed-form expressions for the case of a Hermitian generator times the imaginary unit [\cref{eq:hermitian_1,eq:hermitian_2}].
We illustrate how fermionic rotations may be applied to transformations generated by a sum of terms when combined with Trotterization.
We consider a 5-site Hubbard model with open boundaries  and Hamiltonian defined as:  
\begin{equation}
\mathcal{H} = - J \sum_{i=1}^{4} \sum_{\sigma\in\{\uparrow,\downarrow\}} (a^\dagger_{i + 1,\sigma} a_{i,\sigma} + a^\dagger_{i,\sigma} a_{i+1,\sigma}) + U \sum_{i=1}^{5} n_{i\uparrow} n_{i\downarrow},
\end{equation}
where the hopping parameter $J = 1$ and on-site Coulomb integral $U = 1$.
We focus on the ground state with 3 spin-up and 3 spin-down electrons ($\Psi_\mathrm{gs}$) and follow the dynamics after the sudden removal of a spin-up electron from the first site. Therefore, the initial state of our dynamics is $\ket{\psi(t=0)} \propto a_{1\uparrow} \ket{\Psi_\mathrm{gs}}$.

The average number of spin-up electrons on the first site is then followed by computing the Heisenberg dynamics of the number operator $n_{1\uparrow}(t) = a^\dagger_{1\uparrow}(t) a_{1\uparrow}(t) = e^{i \mathcal{H} t}a^\dagger_{1\uparrow}a_{1\uparrow}e^{-i \mathcal{H} t}$.
Writing the Hamiltonian as a sum of terms, $\mathcal{H} = \sum_{\ell=1}^{N} \mathcal{H}_\ell$, Heisenberg dynamics was implemented by breaking down the total propagation time $t$ into 200 equally spaced steps $\Delta t$, and approximating time propagation of each step with a symmetric second-order Trotter approximation \cite{Suzuki.1985.10.1063/1.526596}, namely,
\begin{equation}
e^{-i \mathcal{H} \Delta t} \approx \prod_{\ell = N}^{1} e^{-i \mathcal{H}_\ell \Delta t / 2}
\prod_{\ell = 1}^{N} e^{-i \mathcal{H}_\ell \Delta t / 2}.
\end{equation}

Panel (a) of \cref{fig:dynamics}  reports the average number of spin up electrons on the first site as a function of time up to $tJ = 25$. After an initial fast relaxation that repopulates site 1 with a spin up electron, the system undergoes complex dynamics.
A comparison of the exact and the Trotterized dynamics is reported in the middle panel of \cref{fig:dynamics}, and shows that the maximum absolute deviation of $\braket{n_{1\uparrow}(t)}$ from the exact dynamics is less than $5 \times 10^{-3}$ ($1.40 \times 10^{-3}$ on average). 
In the lower panel of \cref{fig:dynamics}, we report the importance of each many-body component of the $n_{1\uparrow}(t)$ operator. We use the Euclidean norm ($\| \cdot \|_2$), defined as the square root of the sum of the squared coefficients.
This plot shows that as time progresses, the time-evolved number operator attains a many-body character.
Within the early regime of the fast dynamics ($tJ < 1.5$), $n_{1\uparrow}(t)$ is mostly dominated by one-body operators. However, in the subsequent dynamics, the higher the many-body rank, the larger its importance, up to rank 7, after which the importance decays.
An interesting observation in our applications of the closed-form equations to time evolution is that the Euclidean norm of the one-body operator remains constant despite the value of each operator coefficient changing after each transformation.
This is due to a special property of one-body operators.
Unitary transformations of one-body operators by one-body operators are closed, preserving the Euclidian norm of this component.
In contrast, transformations of one-body operators by two- and higher-body operators by either a Hermitian or anti-Hermitian parameterization of $S$ result in the original one-body operator plus additional terms, $e^{-S} \sqop{p}{q} e^{S} = \sqop{p}{q} + \ldots$, where ``$\ldots$'' stands for operators of rank greater than one, leaving the Euclidian norm of one-body operators unchanged.
The combination of these two effects results in the Euclidean norm of the one-body component being invariant with respect to a general transformation.

\section{Conclusion}
We reported exact, closed-form expressions for certain classes of unitary transformations relevant to a broad range of fermionic quantum many-body problems.
Specifically, we considered unitary transformations of a generic product of fermionic creation and annihilation operators generated by either anti-Hermitian or Hermitian combinations of an arbitrary product of elementary fermionic operators.
Our main result allows us to re-sum the non-terminating commutator expansion, characterizing the above unitary transformations, into a manifestly connected expression involving up to two nested commutators independently of the many-body rank of the transformed operator and generator.

Our study demonstrates the potential and versatility of closed-form fermionic transformations through two illustrative applications.
First, we considered Hamiltonian downfolding based on exact canonical transformations, which allowed for a detailed analysis of three- and higher-body induced interactions in the H\textsubscript{2} molecule at representative geometries along the dissociation curve.
Our findings showed that the two-body components dominate, while induced higher-body terms decay rapidly as a function of many-body rank.
We also compared our method to the DUCC and QLDSRG(2) approaches, which use approximate transformations, and found that maintaining the exactness of the unitary transformation is essential for achieving sub-m$E_\mathrm{h}$ errors across the entire potential energy curve.
Second, we applied the symmetric second-order Trotter approximation to simulate the Heisenberg dynamics in a 5-site Hubbard model. Simulations of the time-evolved number operator allowed us to analyze the evolution of its many-body character and the increasing importance of higher-body terms over time.

This work makes an important first step towards tackling general unitary transformations in fermionic systems, opening new ways to approach many-body problems.
On the numerical side, an important future application of the closed-form expressions reported here is realizing canonical transformations and quantum dynamics in the Heisenberg picture with numerical control of the error introduced by operator truncation.
In addition, the availability of closed-form expressions may be useful in mathematical analyses of operator approximations, formal aspects and emulation of fermionic quantum computation \cite{Bravyi.2002.10.1006/aphy.2002.6254}, and studies of exactly solvable models.

\section*{Acknowledgements}This work is supported by the U.S.\ Department of Energy under Award No.\ DE-SC0019374.

\appendix

\section{PROOF OF RECURSION RELATIONSHIPS}
\label{app:recursion}

In this section, we derive the two key closure relations needed to derive the recursion relations that allow us to resum the BCH commutator expansion.
We start by proving that the operator $A = T - T^\dagger$, with $T$ defined as in \cref{eq:t_def} satisfies the condition:
\begin{equation}
A^3 = -A.	
\end{equation}
Given the nilpotency of individual creation and annihilation operators [$(\sqop{p}{})^2 = (\sqop{}{p})^2 = 0$], $T$ satisfies the nilpotency condition $T^2 = 0$, since the square of at least one second-quantized operator appears in $T^2$.
As a consequence, the square of $A$ is equal to
\begin{equation}
\begin{split}
\label{eq:A_squared}
	A^2  = &  - (T T ^\dagger + T ^\dagger T) \\
	= & - [n_{s_1} n_{s_2} \cdots (1 - n_{t_1})(1 - n_{t_2})\cdots  \\
	& \;\quad + n_{t_1} n_{t_2} \cdots (1 - n_{s_1})(1 - n_{s_2})\cdots ] \\
	=&  -P.
\end{split}
\end{equation}
The operator $P = T T^\dagger + T^\dagger T$ is Hermitian and idempotent, rendering it a projection operator.
Specifically, it is an operator that projects a state onto the domain of operator $A$, implying that $PA = AP = A$. This relation immediately leads to
\begin{equation}
	\label{eq:A_cubed}
	A^3 = -PA = -A,
\end{equation}
completing the proof.\\
Next, we show that $\comm{\comm{\comm{O}{A}}{A}}{A} = - \comm{O}{A}- 3 A\comm{O}{A}A$. By computing the doubly nested commutator, we obtain
\begin{equation}
	\begin{split}
		\comm{\comm{O}{A}}{A} &{}= \comm{OA-AO}{A}\\
		&{}= \comm{OA}{A} - \comm{AO}{A} \\
		&{}= OA^2 - 2AOA + A^2O
	\end{split}
\end{equation}
Using the above result and \cref{eq:A_cubed}, we obtain that the triply nested commutator equals
\begin{equation}
	\begin{split}
		\comm{\comm{\comm{O}{A}}{A}}{A} &{}= \comm{OA^2 - 2AOA + A^2O}{A}\\
		&{}= \comm{OA^2}{A}-2\comm{AOA}{A} + \comm{A^2O}{A}\\
		&{}= OA^3 - 3AOA^2 + 3A^2OA - A^3O\\
		&{}= -OA + AO -3A(OA - AO)A\\
		&{}= -\comm{O}{A} -3A\comm{O}{A}A.
	\end{split}
\end{equation}

\section{PROOF OF COMMUTATOR RELATIONSHIPS}
\label{app:commutator}

In the following, we show that the term $A[O,A]A$ where $O$ is a product of fermionic operators and $A = T - T^\dagger$ is the anti-Hermitian combination of a product of fermionic operators $T$ equals 0 or $[O,A]$.
This result is crucial to re-sum the BCH expansion of the fermionic rotations discussed in this work.
This result is equivalent to showing that the triple commutator satisfies \cref{eq:c3OA}, where $\alpha$ can take only the values 1 or 4, leading to \cref{eq:sto_case1,eq:sto_case2}.

First, we consider two second quantized operators $O$ and $T$ with no common indices.
In this case, $O$ and $A$ either commute or anticommute.
In the commuting case, the fermionic rotation leaves the operator $O$ unchanged.
In the anticommuting case, $A[O,A]A = [O,A] = 2OA$, which corresponds to the case $\alpha = 4$.
An example of this case is $O = \sqop{p_1}{}$ and $T = \sqop{s_1 s_2 s_3}{}$, for which $A[O,A]A = [O,A] = 2 \sqop{p_1 s_1 s_2 s_3}{}$ (here and in the following, we assume indices labeled by different symbols to represent distinct spinorbitals).

When $O$ shares at least one index with $T$, then we consider the full expression $A[O,A]A$.
Upon expanding the commutator and using the definition of $A$ we can rewrite $A[O,A]A = T^\dagger (TOT) - T^\dagger (TOT^\dagger) + T(T^\dagger OT) - T (T^\dagger O T^\dagger) +\ldots$, in terms of four distinct sequences of operators: $TOT$, $TOT^\dagger$, $T^\dagger O T$, and $T^\dagger O T^\dagger$.
Next, we show that, depending on the indices shared by $O$ and $T$/$T^\dagger$, all terms are either simultaneously zero or only one of them is nonzero at a time.
For this analysis, it is convenient to denote the number operator components of $T$ as $T_n$ and the remaining operators as $\tilde{T}$:
\begin{equation}
\label{eq:t_def2}
T = \underbrace{
\nop{r_1} \nop{r_2} \cdots
}_{T_n}
\underbrace{
\sqop{s_1 \cdots s_l}{t_1 \cdots t_k}
}_{\tilde{T}}
= T_n \tilde{T}.
\end{equation}
Consider the term $TOT$, which we write using the definitions in \cref{eq:o_def,eq:t_def2} as:
\begin{equation}
TOT =
\underbrace{
\nop{r_1} \nop{r_2} \cdots  \sqop{s_1 \cdots s_l}{t_1 \cdots t_k}
}_{T}
\underbrace{
\sqop{p_1 p_2 \cdots}{q_1 q_2 \cdots}
}_{O}
\underbrace{
\nop{r_1} \nop{r_2} \cdots 
\sqop{s_1 \cdots s_l}{t_1 \cdots t_k}
}_{T}.
\end{equation}
For $TOT$ to be nonzero, the operator $O$ must satisfy two conditions. First, to avoid the creation/annihilation operators of the two copies of $\tilde{T}$ ($\sqop{s_1 \cdots s_l}{t_1 \cdots t_k}$) from colliding with each other (yielding a zero due to the anticommutator property of identical operators), $O$ must contain the adjoint of this portion of $\tilde{T}$:
\begin{equation}
O = \sqop{t_1 \cdots t_k p_{k+1}\cdots}{s_1 \cdots s_l q_{l+1} \cdots}
= \sqop{t_1 \cdots t_k}{s_1 \cdots s_l} O' =\tilde{T}^\dagger O',
\end{equation}
where $O' = \pm \sqop{p_{k+1}\cdots}{q_{l+1} \cdots}$ contains the remaining operator indices and without losing generality, we assume that the first $k$ upper and $l$ lower indices of $O$ match those of $\tilde{T}^\dagger$.
Note that since $\tilde{T}^\dagger$ and $O'$ share no indices, they either commute or anticommute (and analogously for $\tilde{T}$).
If $O'$ shares indices with $T_n$, those must correspond to number operators, otherwise $TOT = 0$.
When $O$ takes this form, it can be shown that the remaining three sequences $TOT^\dagger$, $T^\dagger O T$, and $T^\dagger O T^\dagger$ are null.
Then it follows that
\begin{equation}
\begin{split}
A[O,A]A & =  T^\dagger TOT - TOT T^\dagger \\
& = T_n \tilde{T}^\dagger \tilde{T} \tilde{T}^\dagger O'T - T\tilde{T}^\dagger O'\tilde{T}  \tilde{T}^\dagger T_n \\
& = T_n \tilde{T}^\dagger O'T - T \tilde{T}^\dagger O' T_n\\
& = OT - T O\\
& = [O,A],
\end{split}
\end{equation}
where we use the identities $T_n^2 = T_n$ and $\tilde{T}^\dagger \tilde{T}\tilde{T}^\dagger=\tilde{T}^\dagger$ and the fact that $OT^\dagger = T^\dagger O = 0$ in the last step.
This corresponds to the case $\alpha = 4$.
As an example, consider the case of $O = \sqop{p_1 p_2}{q_1 q_2}$ and $T = \sqop{q_2}{p_2}$.
In this case, $TOT = \sqop{q_2}{p_2} \sqop{p_1 p_2}{q_1 d} \sqop{q_2}{p_2} = -\sqop{p_1 q_2}{q_1 p_2}$, and it easy to verify that $A[O,A]A = [O,A] =  \sqop{p_1}{q_1} (\nop{p_2} - \nop{q_2})$.
The term $T^\dagger O T^\dagger$ may be handled in a similar way, with the only nonzero contribution arising from operators of the form $O = \tilde{T} O'$, where $\tilde{T}$ and $O'$ share no indices. This also falls under the case $\alpha = 4$.

Next, we examine the case $TOT^\dagger$:
\begin{equation}
	TOT^\dagger = \nop{r_1} \nop{r_2} \cdots  \sqop{s_1 \cdots s_l}{t_1 \cdots t_k} a_{q_1 q_2 \cdots}^{p_1 p_2 \cdots} \sqop{t_1 \cdots t_k}{s_1 \cdots s_l}  \nop{r_1} \nop{r_2} \cdots.
\end{equation}
To avoid the creation/annihilation operators of $O$ from colliding with the ones in $T$/$T^\dagger$, no common indices can exist between $O$ and $\tilde{T}$ and $\tilde{T}^\dagger$, except for number operators with indices in common with the creation part of $\tilde{T}^\dagger$. $O$ can also share indices with the number operator part of $T$, but those should be in the form of number operators to prevent $TOT^\dagger = 0$.
Under these restrictions, then $O$ must have this form
\begin{equation}
O = 
\sqop{t_{\mu_1} \cdots t_{\mu_m} p_{m+1}\cdots}{t_{\mu_1} \cdots t_{\mu_m} q_{m+1} \cdots}
= \sqop{t_{\mu_1} \cdots t_{\mu_m}}{t_{\mu_1} \cdots t_{\mu_m}} O' =O_n O',
\end{equation}
where $\{t_{\mu_1},\ldots,t_{\mu_m}\}$ is a subset of the indices $\{t_1, \ldots, t_k\}$ and for convenience, we assume that these indices appear first in the definition of $O$.
The quantity $O_n$ is a number operator such that $T O_n = T$ and $O_n  T^\dagger =  T^\dagger$, while $O'$ is the component of $O$ that commutes or anticommutes with $T$ and $T^\dagger$.
Note that since $O$ and $T$ must share indices, if $O_n$ is the identity, then $O'$ must have number operators in common with $T_n$. In this case, $O$ and $T$ either commute or anticommute, reducing to the case of no shared indices.
If, instead, $O_n$ contains at least one number operator, then $O_n O' = O' O_n$ since the indices of the operators in $O_n$ are distinct from those of $O'$.
From these considerations, it is straightforward to evaluate $A[O,A]A$:
\begin{equation}
\begin{split}
A[O,A]A & =  -TOT^\dagger T - T^\dagger T O T^\dagger \\
& = -T O_n O' T^\dagger T - T^\dagger T O_n O' T^\dagger\\
& = -T T^\dagger T O' - O' T^\dagger T T^\dagger\\
& = -T O' - O' T^\dagger\\
& = -T O_n O' - O' O_n T^\dagger\\
& = [O,A],
\end{split}
\end{equation}
where in the last step, we use the fact that $OT = T^\dagger O = 0$.
This result also corresponds to the case $\alpha = 4$.
As an example, consider the case of $O = \sqop{p_1 p_2}{p_1 q_1}$ and $T = \sqop{s_1}{p_1}$.
In this case, $TOT^\dagger = \sqop{s_1}{p_1} \sqop{p_1 p_2}{p_1 q_1} \sqop{p_1}{s_1} = \sqop{p_2}{q_1} \nop{s_1} (1 - \nop{p_1})$, and after evaluating commutators it is possible to verify that $A[O,A]A = [O,A] = \sqop{p_1 p_2}{q_1 s_1} +\sqop{p_2 s_1}{p_1 q_1}$.

Following the same procedure, one may prove that when $T^\dagger O T \neq 0$ also leads to $\alpha = 4$.
Lastly, when $O$ is such that all four combinations $TOT$, $TOT^\dagger$, $T^\dagger O T$, $T^\dagger O T^\dagger$ are null, then $A[O,A]A = 0$ and we recover the case $\alpha = 1$.
An example of this case is given by $O = \sqop{p_1 p_2}{p_1 q_1}$ and $T = \sqop{s_1}{p_2}$, which leads to $A[O,A]A = 0$ while $[O,A] = -\sqop{p_1 s_1}{p_1 q_1}$.
This concludes our proof for the anti-Hermitian combination of the generator $T$.

The proof for the Hermitian case [\cref{eq:hermitian_1,eq:hermitian_2}] follows along the same lines, noting that if $T$ is not a number operator, $H^3 = H$ and $[[[O,H],H],H] = [O,H] - 3H[O,H]H$. Using a similar strategy followed for the anti-Hermitian case, it can be shown that $H[O,H]H$ is either $-[O,H]$ or 0. The case of $T$ being a number operator can be easily considered separately.

\newpage


%

\end{document}